\documentclass[%
 prd,
cp,  
 amsmath,amssymb,
 reprint,%
]{revtex4-2}

\usepackage{graphicx}
\usepackage{dcolumn}
\usepackage{bm}
\usepackage{hyperref}  

\usepackage[utf8]{inputenc}
\usepackage[T1]{fontenc}
\usepackage{mathptmx} 
\usepackage{nameref}
\usepackage{subcaption}
\usepackage{amssymb}

\usepackage{color} 

\begin{document}

\title{Observational Constraints on the Maximum Masses of White Dwarfs, Neutron Stars, and Exotic Stars in Non-Minimal Derivative Coupling Gravity}

    \author{M. D. Danarianto$^{a,b}$}
	\email{m.dio.danarianto@brin.go.id}
	\author{I. Prasetyo$^{c,d}$}
	\email{ilham.prasetyo@sampoernauniversity.ac.id}
	\author{A. Suroso$^{e}$}
	\email{agussuroso@fi.itb.ac.id}
	\author{B. E. Gunara$^{e}$}
	\email{bobby@itb.ac.id}
	\author{A. Sulaksono$^b$}
	\email{anto.sulaksono@sci.ui.ac.id}

    \affiliation{$^a$Research Center for Computing, National Research and Innovation Agency (BRIN), Bandung 40173, Indonesia}
    \affiliation{$^b$Departemen Fisika, FMIPA Universitas Indonesia, Kampus UI, Depok 16424, Indonesia}
	\affiliation{$^c$Department of General Education, Faculty of Art and Sciences, Sampoerna University, Jakarta, Indonesia}
	\affiliation{$^d$IoT and Physics Lab, Sampoerna University, Jakarta, 12780, Indonesia}
	\affiliation{$^e$Theoretical High Energy Physics Research Division, Institut Teknologi Bandung, Jl. Ganesha 10 Bandung 40132, Indonesia}

\date{\today} 

\begin{abstract}
The advancement of astronomical observations opens the possibility of testing our current understanding of gravitational theory in the strong-field regime and probing any deviation from general relativity.  We explore to what extent compact stars predicted by non-minimal derivative coupling (NMDC) gravity theory agree with observed data. We investigate white dwarfs (WDs), neutron stars (NSs), and quark stars (QSs) mass and radius in various values of constant scalar $|Q_{\infty}|$ at coupling strength of $\eta=\pm1$. This study focuses on the astrophysical impacts of altering maximum masses by values of $|Q_{\infty}|$ and $\eta$. From an observational point of view, we found that WD stars are consistent with ultra-cold WD data at $|Q_{\infty}| \lesssim 0.2$. We also found that QS has a similar impact of mass-radius to NS, where the modification is more significant at higher (central) density. For NS and QS EoSs, the value $|Q_{\infty}|$ strongly alters the critical mass and might eliminate the $M-\rho_c$ turning point in the negative $\eta$ case. In that case, the sufficiently large $|Q_{\infty}|$ could predict $M>2.6 M_\odot$ NS and QS, i.e., larger than GW190814 secondary counterpart. We suggest that the lower mass gap in the gravitational wave and x-ray binary mass population data might restrict the theory's $|Q_{\infty}|$.
\end{abstract}

\maketitle
\section{\label{sec:introduction} Introduction}

General Relativity (GR) theory is one of the pillars of physics because almost all fundamental aspects of GR have been tested with flying colors through numerous observations and experiments from the ground to the cosmic scale. GR has passed all experimental tests in the Solar System and binary pulsars\cite{Will:2014kxa}. Even recently, the surprising prediction of GR, black holes (BHs), has been confirmed by detecting gravitational waves due to a merger of two black holes\cite{LIGOScientific:2020iuh}.  The multimessenger observation of GW170817 \cite{LIGOScientific:2017ync,LIGOScientific:2017vwq} and the imaging of accretion disks in M87, and Sgr A* can be explained by GR theory\cite{EventHorizonTelescope:2019dse,EventHorizonTelescope:2022wkp}. Nevertheless, despite this success, GR still cannot satisfactorily explain several gravity issues related to the process at the ultraviolet (UV) and infrared (IR) regimes. For example, cosmology has tried to solve the unknown nature of dark matter and dark energy and understand the space-time singularity~\cite{Berti:2015itd}. There are also some issues related to compact objects. For example, the issue of the maximum allowed mass of horizonless compact objects exceeded GR predictions. Please see the corresponding discussions for neutron stars (NSs) in Refs.~\cite{Sun:2023glq,Linares:2018ppq,Margalit:2017dij}, while for quark stars (QSs), please see Refs.~\cite{Goswami:2023tps,Yang:2023haz}, and for white dwarfs (WDs), please see Refs~\cite{Astashenok:2022kfj,Astashenok:2022uhw,Sarmah:2021ule} and the related references therein. The issue related to the nature of ultra-compact low-mass compact star HESS J1731-347 has recently become a hot discussion in the literature. For example, the corresponding issue is discussed in Refs. \cite{Sagun:2023rzp,Kubis:2023gxa,Li:2023vso,Horvath:2023uwl,Rather:2023tly,Oikonomou:2023otn}.  It was reported n Ref.\cite{LIGOScientific:2020zkf}
that a binary (a BH and an unidentified compact object) merger through the GW190814 event. The exact nature of the unidentified object is also discussed quite recently. For example, the corresponding discussion can be found in Refs. \cite{Prasetyo:2021kfx,Rahmansyah:2021gzt} and please see the references therein. Furthermore, it is well known that the equation of state (EoS) of horizonless compact objects is still being determined. Please see the corresponding discussion in Refs.\cite{Malik:2024nva,Landry:2020vaw,Danarianto:2023rff,Horvath:2023uwl,Chu:2023rty}.  Recent work has also shown that gravity could influence the EoS of these compact objects~\cite{Wojnar:2022dvo}.  Therefore, gravitational theories other than GR are expected to solve this strong-gravity regime while consistent with the GR predictions in the intermediate energy regime.

On the other hand, GR has been tightly constrained in a vacuum, leaving little room for modification in the absence of matter. For example, the gravitational wave speed measured by GW170817 ruled out several classes of modified gravity \cite{Kase2019}. This turns our attention to the theories that modify gravity only in the interior of the objects while reducing back to the GR metric for the exterior solution. Several modified gravity has this feature, for example, Eddington-inspired Born Infeld \cite{Banados2010} and type-I of minimally modified gravity \cite{Aoki2018}. However, models like Eddington-inspired Born Infeld have problems such as the singularity at the stellar surface.

One of the possible modifications to GR that satisfy the later phenomenological stringent constraints and be theoretically viable like free of ``Ostrogradski ghost'' is by addition of a new dynamical degree of freedom in the form of a scalar field coupled to gravity~\cite{Cisterna:2015yla,Maselli:2016gxk}.  Several studies conclude that a modification of GR obtained from the most general action with a single scalar degree of freedom with the above requirements corresponds to the scalar-tensor theory formulated by Horndeski.
Hordenski's theory of action yields an equation of motion containing, at most, second-order derivatives, and the theory is invariant under Galilean shift symmetry in flat to-curve space times.  Fab Four is Hordenski's theory with additional restrictions that allow for dynamical self-tuning of quantum vacuum energy. Please see Refs. \cite{Maselli:2016gxk,charmousis2012general} and the references therein for detailed discussion. If we select only the George and John terms from the Fab Four theory, the theory becomes the one known as non-minimal derivative coupling (NMDC) gravity theory~\cite{Cisterna:2015yla,Maselli:2016gxk,charmousis2012general}. The theory has a  $G^{ab}\nabla_a\Phi\nabla_b\Phi$ in action, where $G^{ab}$ is the Einstein tensor and $\Phi$ is a scalar field. Note that in the NMDC theory, a small but nonzero dynamic cosmological constant naturally appears, which triggers the acceleration of cosmological expansion in late times. Note that it is known that the predicted cosmological constant from the current physics is much larger than the observed cosmological constant from various astronomical observations. There is also another feature that the NMDC theory exhibits. The theory yields some black-hole (BH) solutions that have a scalar hair \cite{Minamitsuji:2013ura,Babichev:2013cya,rinaldi2012black}. However, the model has other attractive BH solutions with similarly intriguing consequences to the scalar field's current density vector~\cite{momeni2016spherically}. These BH's solutions evade the no-hair theorem, even when the regularity conditions are satisfied. These regularity conditions originated from a no-hair theorem for Galileon~\cite{hui2013no}, then altered by the authors of \cite{Babichev:2013cya} to evade the no-hair theorem, producing solutions called \emph{stealth solutions}. Furthermore, Cisterna, Delsate, and Rinaldi \cite{Cisterna:2015yla} studied this model for static NSs. The odd-parity perturbation aspect is studied by Cisterna, Cruz, Delsate, and Saavedra~\cite{cisterna2015nonminimal}. The properties of slowly-rotating NS within NMDC are studied by Cisterna, Delsate, Ducobu, and Rinaldi~\cite{Cisterna:2016vdx}. It is worth noting that BSs and NSs within Hordenski and beyond Hordenski theories were studied in Refs.\cite{Maselli:2016gxk,Maselli:2015yva,Bakopoulos:2023sdm,Bakopoulos:2022csr,Barranco:2021auj,Chagoya:2018lmv,Silva:2016smx}.

In this paper, we systematically investigate the static properties of compact stars, i.e., WDs, NSs, and QSs predictions of NMDC theory~\cite{Cisterna:2015yla}, and the results are compared to the current observational data. The purpose of this work is to test the viability of NMDC theory predictions for compact star static properties. We also investigate the impacts of various NMDC parameter $\eta$ and ansatz parameter $Q_\infty$ variations on compact stars' mass-radius relation. The latter related to the following questions: Are the restriction of $\eta=\pm1$ and $|Q_{\infty}| \leq 0$ compatible with the masses and radii of compact stars? How much of the $|Q_{\infty}|$ variation does alter the (maximum) mass and radius of these objects within various EoSs and their prediction's compatibility with observational data?

This paper has been organized as follows: Section \ref{sec:gravity} briefly reviews the NMDC theory and the stellar structure of compact stars within this theory. In section \ref{sec:results}, we provide the numerical results of the stellar structure of compact stars for WD, NS, and QS cases within NMDC theory. Finally, the concluding remarks are given in Section ~\ref{conclu}.

\section{Stellar structure in non-minimal derivative coupling}
\label{sec:gravity}

In this section, we review the derivation of the crucial equations of NMDC theory for numerical calculation. The latter will be discussed in more detail in the next section. Our work starts from a scalar-tensor theory by Cisterna {\it et al.} \cite{Cisterna:2015yla}. The action is given as
\begin{equation}
	S=\int \sqrt{-g} d^4x \left[
	\kappa(R-2\Lambda) -\frac{1}{2} (\alpha g^{ab} - \eta G^{ab})\nabla_a\Phi \nabla_b\Phi
	\right] + S_m,
\end{equation}
with $\kappa=1/(16\pi G)$ and $G$ is the Newton's constant. $\Lambda$ is the cosmological constant and $g$ is the determinant of the spacetime metric $g_{ab}$. The kinetic term of the scalar field $\Phi$ is multiplied by its coupling constant $\alpha$. This term is called the minimal derivative term, while the non-minimal derivative term is multiplied by $\eta$. $S_m$ denotes action from matter contribution.
The equations of motion can be obtained by variational principle:
\begin{subequations}
	\begin{align}
		\nabla_a T^{ab}=0 \text{~with~}& T_{ab}=\frac{\delta S_m}{\delta g^{ab}},\\
		\nabla_a J^a=0 \text{~with~}& J^a=(\alpha g^{ab}-\eta G^{ab})\nabla_b\Phi,\\
		G_{ab}+\Lambda g_{ab} - H_{ab} &= (2\kappa)^{-1} T_{ab},
	\end{align}
with
\begin{equation}
	H_{ab}=\sum_{n=1}^2 \frac{\alpha}{2\kappa} H^{(n)}_{ab}+ \sum_{n=3}^{11} \frac{\eta}{2\kappa} H^{(n)}_{ab}.
\end{equation}
\end{subequations}
The minimal derivative coupling term gives two terms
\begin{subequations}
\begin{align}
	H^{(1)}_{ab}&=\nabla_a\Phi \nabla_b\Phi,\\
	H^{(2)}_{ab}&=-\frac{1}{2} g_{ab}\nabla^c\Phi \nabla_c\Phi,
\end{align}
\end{subequations}
while the non-minimal derivative coupling term gives nine terms
\begin{subequations}
\begin{align}
	H^{(3)}_{ab}&=
	\frac{1}{2}R \nabla_a\Phi \nabla_b\Phi, \\
	H^{(4)}_{ab}&=
	-\nabla^c\Phi(\nabla_a\Phi R_{bc} + \nabla_b\Phi R_{ac}), \\
	H^{(5)}_{ab}&=
	-\nabla^c\Phi \nabla^d\Phi R_{cadb},\\
	H^{(6)}_{ab}&=
	-\nabla_a\nabla^c\Phi \nabla_b\nabla_c\Phi, \\
	H^{(7)}_{ab}&=
	\frac{1}{2}g_{ab}\nabla^c\nabla^d\Phi \nabla_c\nabla_d\Phi, \\
	H^{(8)}_{ab}&=
	-\frac{1}{2}g_{ab}\left(\nabla_c\nabla^c\Phi\right)^2,\\
	H^{(9)}_{ab}&=
	\nabla_c\nabla^c\Phi \nabla_a\nabla_b\Phi, \\
	H^{(10)}_{ab}&=
	\frac{1}{2}g_{ab}\nabla^c\Phi \nabla_c\Phi G_{ab}, \\
	H^{(11)}_{ab}&=
	g_{ab} R_{cd}\nabla^c\Phi \nabla^d\Phi.
\end{align}
\end{subequations}
The line element that we use is
\begin{equation}
	ds^2= - b(r) dt^2 + {1\over f(r)} dr^2 + r^2 d\theta^2 + r^2 \sin^2\theta d\varphi^2 .
\end{equation}
We consider the case of no cosmological constant and only non-minimal derivative coupling $\Lambda=0=\alpha$. Following Ref. \cite{Babichev:2013cya}, we choose $J^r=0$ to maintain the regularity of scalar current at all $r$ in $0\leq r <\infty$. This choice came from \cite{Hui:2012qt}, whose authors proved that $J^\theta=J^\varphi=J^t=0$ given the space-time is static and spherically symmetric. Because we intend to discuss stars with ideal fluid, whose mass density and pressure are $\rho$ and $P$ respectively, we used
\begin{equation}
	T^a_{~b}=\text{diag}[-\rho (r),P(r),P(r),P(r)].
\end{equation}
The scalar field ansatz is\cite{Babichev:2013cya}
\begin{equation}
	\Phi=Qt+F(r) ,
\end{equation}
where $Q$ is a constant parameter called ansatz parameter. Alongside this ansatz is the constraint $J^r=0$ and $\partial\Phi/\partial r \neq 0$ to ensure non-divergent current density squared $J^aJ_a$. The ansatz also gives rise to the so-called \emph{stealth solutions}, which indicates a nontrivial scalar hair. This ansatz is further studied in the case of rotating BHs by Bakopoulos, Charmousis, and Lecoeur~\cite{bakopoulos2022compact}.

Through a long calculation, one can arrive at the following results. From the conservation of the stress-energy tensor, we obtain (with $f'=df/dr$)
\begin{subequations}
\begin{equation}
	\label{eq:dpdr}
	P' = -\frac{b' (P+\rho)}{2 b},
\end{equation}
while from $J^r=0$ we have
\begin{equation}
	\label{eq:dbdr}
	b' = -\frac{b (f-1)}{r f}.
\end{equation}
Substituting this into $rr$ component in the modified Einstein Field Equation, we have
\begin{equation}
	F'= \sqrt{\frac{r^2 b P-\eta  Q^2 f+\eta  Q^2}{\eta b f}}.\label{Eq:EOMunused}
\end{equation}
Lastly, substituting these equations into $tt$ component in the modified Einstein Field Equation (EFE) gives us
\begin{align}
	f'(r)&= \frac{3 \eta  Q^2 (f(r)-1) f(r)-b(r) A}{r \left(b(r) \left(4 \kappa +r^2 P(r)\right)-3 \eta  Q^2 f(r)\right)}, \label{eq:f'}\\
	\label{Eq:EOMA}
	A&=f(r) \left(4 \kappa +6 r^2 P(r)+r^2 \rho (r)\right)-4 \kappa +r^2 \rho (r).
\end{align}
\label{Eq:EOM}
\end{subequations}

For the numerical calculation, we integrate Eqs. (\ref{eq:dpdr}-\ref{Eq:EOMA}) starts from the center of a star to its surface. To determine central boundary conditions, we use expansion around $r=r_c\to 0$ to all of the functions:
\begin{subequations}
\begin{align}
	P(r_c) &= P_c + P_1 r_c + P_2 r_c^2 + \ldots,\\
	\rho(r_c) &= \rho(P_c) = \rho_c + \rho_1 r_c + \rho_2 r_c^2 + \ldots,\\
	b(r_c) &= b_c + b_1 r_c + b_2 r_c^2 + \ldots,\\
	f(r_c) &= 1 + f_1 r_c + f_2 r_c^2 + \ldots,\\
	F'(r_c)^2 &= C_0 + C_1 r_c + C_2 r_c^2 + \ldots.
\end{align}
\end{subequations}
Substituting into the equations \eqref{Eq:EOM} gives us the boundary of (only up to one correction term)
\begin{subequations}
\begin{align}
	P(r_c) &= P_c + \frac{b_c^2(P_c+\rho_c)(3P_c-\rho_c)}{6(3Q^2\eta-4\kappa b_c^2)} r_c^2,\label{eq:Pcenter}\\
	\rho(r_c) &= \rho_c,\\
	b(r_c) &= b_c,\\
	f(r_c) &= 1,\\
	F'(r_c)^2 &= \left( \frac{P_c}{\eta} - \frac{2Q^2(3P_c-\rho_c)}{3(3Q^2\eta-4\kappa b_c^2)} \right) r_c^2.\label{eq:Fprimesquaredcenter}
\end{align}
\label{eq:centerconditions}
\end{subequations}
Because pressure structure should be decreasing and go to zero in vacuum, the second term in  Eq. \eqref{eq:Pcenter} should be less than zero and always decreasing. From this, we have bound of $\frac{\eta Q^2}{b_c^2} < \frac{4\kappa}{3}$. Also, because $F$ is a real function, we require a non-negative value to the right-hand side of Eq. \eqref{eq:Fprimesquaredcenter}. Combining them gives us the upper and lower restriction for both $\eta$ and $Q$:
\begin{equation}
	\frac{12 P_c \kappa}{(2\rho_c-3P_c)} < \frac{\eta Q^2}{b_c^2} < \frac{4\kappa}{3}. \label{eq:ineq}
\end{equation}
We also set $\eta=\pm 1$ because on Eq. \eqref{Eq:EOM} (except for Eq. \eqref{Eq:EOMunused}), $\eta$ always multiplied with $Q^2$.

In performing the integration, we start the calculation by setting the initial value at $r=r_c$ using Eqs. \eqref{eq:centerconditions}, using a single $P_c$ value as the input. At this point, $b_c=b_{c\text{(old)}}=1$ and $Q=Q_{\infty}=Q_{\text{(old)}}$. This calculation is continued until the pressure reaches zero at surface, $r=R$. However, this might gives us different metric solution at the surface, $b(R)$ and $f(R)$
\begin{equation}
b(R) = (1-2GM/R)/b_{\text{corr}} = f(R)/b_{\text{corr}},
\end{equation}
with $b_{\text{corr}}\neq 1$ as the metric correction ratio. Because $b(R)$ and $f(R)$ must coincide in vacuum, we need to repeat the calculation with the same $P_c$ by rescaling both $b_c$ and $Q$ before recalculating the again structure. Applying scale invariance characteristics of $b'$, now we can rescale $b_c$ using
\begin{equation}
	b_{c\text{(new)}}=b_{c\text{(old)}} b_{\text{corr}}.
\end{equation}
This in turn redefine $t\to t/\sqrt{b_{\text{corr}}}$ and thus we also rescale $Q$ with
\begin{equation}
	Q_{\text{(new)}}=\frac{Q_{\text{(old)}}}{\sqrt{b_{\text{corr}}}}.
\end{equation}
We repeat above scaling steps iteratively until $|1-b_{\text{corr}}|\leq 10^{-3}$, which now give us the final value of $Q_{\text{(new)}}$. We should add that $Q_{\text{(new)}}=Q$ (with $b_c=1$) still have to satisfy the inequality \eqref{eq:ineq}. The key point is guessing the $Q_{\text{guess}}$ value in the initial $Q$ value $Q_{\infty}=\frac{4\kappa}{3} Q_{\text{(guess)}}$. It turns out that we only need to start the iteration from a small $Q_{\text{(guess)}}$ value because $Q_{\text{(new)}}$ tends to increase its value (see Fig. 3 in Ref. \cite{Cisterna:2015yla}).

This procedure is the reverse method of what was given by Ref. \cite{Cisterna:2015yla} where they start from $Q$ to get $Q_\infty$ with $Q_{\text{(new)}}=Q_{\text{(old)}}\sqrt{b_{\text{corr}}}$. Here, we start from $Q_\infty$ to get $Q$. We chose this approach to ensure the plot was shown with values of $Q_\infty$ rather than $Q$ as in \cite{Cisterna:2015yla}.

\section{Numerical results and discussion}
\label{sec:results}
Using the algorithm explained above, we calculate numerically the structure of the white dwarf, strange star, and neutron star using the EoS described in this section. For this purpose, we implement the numerical integration procedures, namely the $4^{th}$-order Runge-Kutta method. The calculated mass-radius are then qualitatively compared to the relevant observational measurements with the types of the stars. This section will discuss the numerical results and their implications for the current understanding of compact stars by confronting the results with observational data. We also compare them with the GR solution for each EoS as a benchmark of the analysis. Here, the value of $|Q_{\infty}|$ are normalized by the factor of $\sqrt{4\kappa/3}$.

\subsection{White dwarf: Hamada-Salpeter EOS}
\textit{White dwarf equation of state.} The white dwarf structure is supported mainly by degeneracy pressure between electrons as described by Chandrasekhar EoS \cite{Chandrasekhar:1931ftj} as the relativistic Fermi gas at zero temperature. Here, we apply the refinement of Chandrasekhar EoS, namely Hamada-Salpeter (HS) \cite{1961ApJ...134..683H}, which accounts for electrostatic (Coulomb) interaction between electrons and ions, Thomas-Fermi, electron exchange, and electron correlation in addition to zero-temperature Fermi gas. In general, this EoS would reduce the overall critical mass by about 5\% below the Chandrasekhar EoS, i.e., due to the nature of the attractive behavior of the Coulomb interaction term.

It has also been known that the critical mass is affected by the EoS and the relativistic effects of gravity. However, the difference is slight in this scale and could be negligible except around the critical mass. For example, the mass of WD between Newtonian and general relativity has only about 1\%  difference in maximum mass \cite{Carvalho:2017kyk,Althaus:2022vwr,Althaus:2023kgf} and could be negligible below $1.20M_\odot$. Therefore, this theory's entire relativistic WD structure would also resemble their (modified-)Newtonian one in this scale.

\begin{figure}
	\centering
	\begin{subfigure}[b]{0.5\textwidth}
		\includegraphics[width=\linewidth]{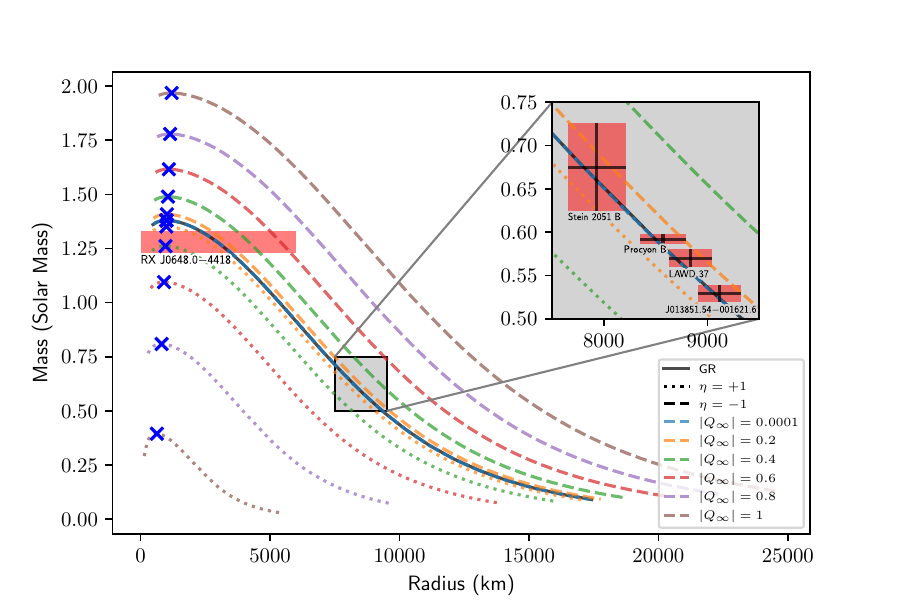}
		\label{fig:mvsrwd}
	\end{subfigure}
	\centering
	\begin{subfigure}[b]{0.5\textwidth}
		\includegraphics[width=\linewidth]{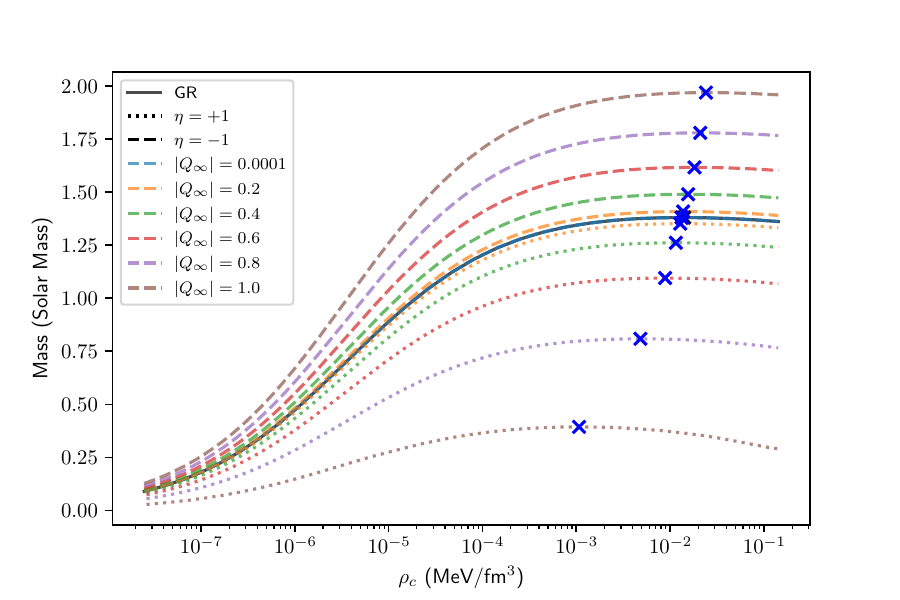}
		\label{fig:mvsrhowd}
	\end{subfigure}
	\caption{The mass-radius curve (upper) and the mass-central density curve (lower) of Hamada-Salpeter equation of state for both $\eta=+1$ (dotted lines) and $\eta=-1$ (striped lines) cases, plotted for each value of $|Q_{\infty}|$ (denoted by different color). A general relativistic case denoted by a solid black line coincides with a small $|Q_{\infty}|$ case. The blue `x' marks indicate each curve's maximum (critical) mass. As the observational comparison, mass-radius curves are overlaid with measured data from X-ray WD binary \cite{Mereghetti:2009pjo} to massive WD and ultracold WDs from Refs. \cite{10.1111/j.1365-2966.2012.21773.x,10.1093/mnras/stac3532,Sahu:2017ksz,Bond2015} to represents the low temperature WD. For $\eta=+1$ case, the brown line labeled by $|Q_{\infty}| = 1$ is actually calculated with $|Q_{\infty}| = 0.95$ due to the existence of positive pressure gradient as $|Q_{\infty}| \rightarrow 1$ similar to Eq. \ref{eq:ineq} boundary.}
	\label{fig:mvsr_mvsrhowd}
\end{figure}

\begin{figure}
	\centering
	\includegraphics[width=\linewidth]{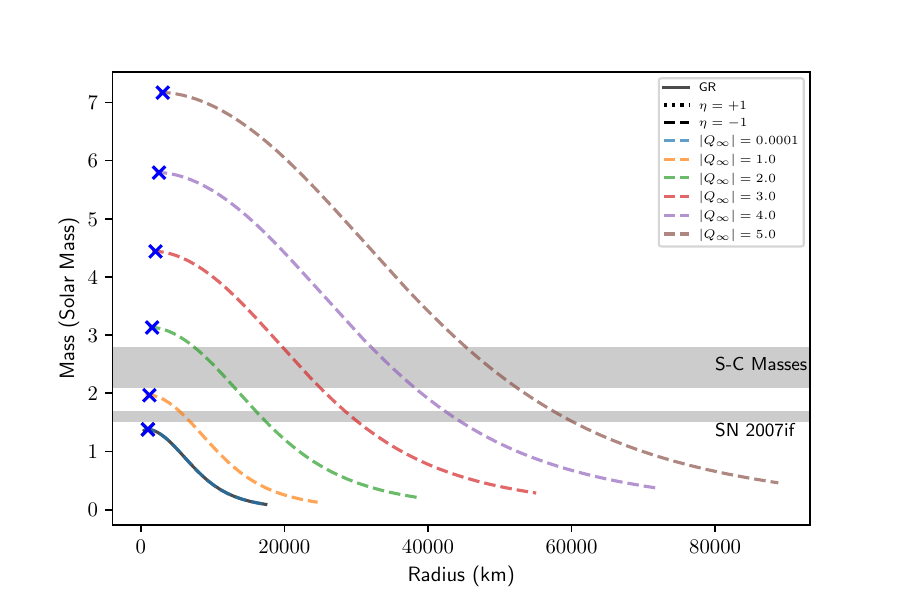}
	\caption{The mass-radius curve for the larger $|Q_{\infty}|$ for $\eta=-1$ case. The lower gray shade shows the estimated mass of SN 2007if \cite{Scalzo2010}, while the upper gray shade shows the typical super-Chandrasekhar supernova mass range \cite{SNLS:2006ics,Hicken2007,Yamanaka2009,Silverman2011,Taubenberger2011}.}
	\label{fig:massrad_HSlargeQ}
\end{figure}
\textit{Numerical result.} The numerical solution for the fully relativistic star using HS EoS is presented in Fig. \ref{fig:mvsr_mvsrhowd}. It is apparent that both $|Q_{\infty}|$ and $\eta$ have a role in adjusting the M-R curve of the white dwarf. The positive value of $\eta$ tends to reduce WDs' mass and radius, while the negative one increases their mass and radius. The value of $|Q_{\infty}|$ amplifies the above effects, giving significant corrections relative to GR within the tenth order. The lowest maximum mass solution on $\eta=+1$ is limited by the positive pressure gradient value at around $|Q_{\infty}|\rightarrow1$ stars (similar to \ref{eq:ineq}). At $|Q_{\infty}| = 0$, the mass, radius, and central pressure relation resembles the value of the GR one. This finding is consistent with the Newtonian limit of this theory studied by \cite{Maselli:2016gxk}, which shows that the theory would approach standard Newtonian gravity as $Q_\infty \rightarrow 0$.

Despite its zero temperature approach, HS EoS is adequate to explain massive WD (i.e., this EoS can set a well-known maximum stable mass of WD $\sim 1.4 M_\odot$) because the temperature effect is less prominent in massive and low-temperature WD as discussed in Refs. \cite{Nunes:2021jdt,Saltas:2018mxc,Danarianto:2023rff}. This critical maximum value has been used as the weighing stone to define the type-1a supernova luminosity as the standard candle in cosmology. We define the similar critical mass as the maximum turning point at the $M-R-\rho_c$ curve (i.e., Fig. \ref{fig:mvsr_mvsrhowd}). This is related to the stability against the gravitational collapse of $dM/d\rho_c > 0$. However, it should be noted that this criterion does not guarantee overall stability since there are minor corrections due to the effects of modified gravity that should be considered.

Based on our numerical result, the value of $\eta$ and $|Q_{\infty}|$ significantly controls the critical mass $M_{crit}$. Interestingly, these values lay at a similar radius around $R_{crit}\approx10^3$ km. This also applies to the smallest WD given by $\eta=+1$ to the highly massive critical mass at $|Q_{\infty}|\gg1$ in $\eta=-1$ (see Fig. \ref{fig:massrad_HSlargeQ}).

\textit{Observational consequences.} It is interesting to see different bounds from different measurements. The measurements from low-temperature WD give bound within the range of intermediate-mass WD ($M\sim 0.50-0.75 M_\odot$). Within this range, temperature has a significant effect on the radius. Therefore, we select several low-temperature WDs ($T_{eff} \lesssim 10^4 K$) with high precision to represent the masses and radii of WD with minimum temperature effect. The above data measured from multiple combinations of methods \cite{10.1111/j.1365-2966.2012.21773.x,10.1093/mnras/stac3532,Sahu:2017ksz,Bond2015}, i.e., astrometric microlensing, spectroscopy, parallax measurements. 

In principle, one may do statistical analysis to determine parameter probability distribution from mass-radius data as in \cite{Saltas:2018mxc,Danarianto:2023rff}. However, the data we present here have approximately a similar range of errors. Therefore, for rough analysis, low-temperature WD gives the bound of $|Q_{\infty}\lesssim0.2|$ for both $\eta=\pm 1$ cases.

We investigate the bound around critical mass using massive WD measurement. We make use of the observed mass-radius of ultramassive  WD RX J0648.0-4418 \cite{Mereghetti:2009pjo} (see Fig. \ref{fig:mvsr_mvsrhowd}), measured from the X-ray eclipse of peculiar WD-subdwarf binary. Note that for high mass WD, the radius is less sensitive to temperature \cite{Althaus:2022vwr,Althaus:2023kgf} and, therefore, could be represented by relatively hot WD. 

The mass of RX J0648.0-4418 gives a limit on $|Q_{\infty}|$ for $\eta=+1$ case via lowest critical mass. The value of $|Q_{\infty}|$ should be $\lesssim 0.4$  in order to be able to reach the mass of RX J0648.0-4418. The upper limit of RX J0648.0-4418 radius (from the rotational stability of measured spin period) also gives a strong upper bound of $|Q_{\infty}|$ for $\eta=-1$ case, $|Q_{\infty}\lesssim 0.6|$. The minimum critical mass also gives bound for $\eta=-1$ case, with $|Q_{\infty}\lesssim0.2-0.3|$.

The measurements above are model-independent; mass and radius are measured separately without dependence on structural models. These methods ensure that the measurements are derived without assuming the gravity structure.

It is interesting to note that at $|Q_{\infty}|>1$ for $\eta=-1$, the mass might exceed the super-Chandrasekhar supernova masses (see Fig. \ref{fig:massrad_HSlargeQ}). The mass would also reach the typical stellar-mass blackhole range $M\gtrapprox5M_\odot$ for $|Q_{\infty}|\gtrapprox5$. However, applying significant $|Q_{\infty}|$ would also increase the mass and radius at lower (central) pressure WD, making it impossible to explain typical WD simultaneously.

\textit{Note on the scalar field in WD solution.} As discussed in Ref. \cite{Cisterna:2015yla}, the problem with the $\eta=-1$ case is the existence of a negative value of $F'^2$ inside the stellar structure solution. It is known that the negative value of $F'^2$ is more dominant on the lower value of $|Q_{\infty}|$. Here, we also found a similar pathology on HS EoS around a small value of $|Q_{\infty}|\lesssim0.1-0.2$ (see Fig. \ref{fig:F2_WD}). Note that this range is inside the observational WD bounds and coincides with the GR M-R curve.

\subsection{Neutron star: Parameterized high-density EOS}

\textit{Neutron star equation of state.} Since there is no general agreement on the EoS in high density, several EoS options represent a neutron star matter. One helpful method to represent various EoS introduced by Ref. \cite{Read:2008iy} is by fitting numbers of nuclear matter EoS into the parameterized piecewise polytropic EoS form. In this approach, the EoS is approximated by dividing $\rho$ into numbers of polytropic EoS, i.e.,
 \begin{equation}
 	P=K_i \rho^{\Gamma_i},
 \end{equation}
 at $\rho_{i-1}<\rho<\rho_i$. The EoS is divided into three zones of density separated by $\rho_1=10^{14.7}$ g/cm$^3$ and $\rho_2=10^{15.0}$ g/cm$^3$. A realistic nuclear matter EoS is then fitted and represented by the value of $P_1, \Gamma_1,\Gamma_2,$ and $\Gamma_3$. If these values are in hand, we can calculate the complete piecewise polytropic after calculating $K_i$ by considering continuous EoS via
  \begin{equation}
  	K_{i+1} = \frac{P(\rho_i)}{\rho_i^{\Gamma_{i+1}}}.
  \end{equation}

 We selected three samples of EoS representing the maximum NS mass in GR case for the lowest (PAL6, $M_{max,GR} \sim 1.47 M_\odot$), intermediate (SLy, $M_{max,GR} \sim 2.0 M_\odot$), and the largest (MS1, $M_{max,GR} \sim 2.8 M_\odot$) from parameterized piecewise polytropic EoS from Ref. \cite{Read:2008iy} as the approximation to estimate the structure of well-known neutron star EoS.
 We apply SLy EoS \cite{Douchin:2001sv} for low-density (crust) regions.
 
\begin{figure}
	\centering
	\includegraphics[width=\linewidth]{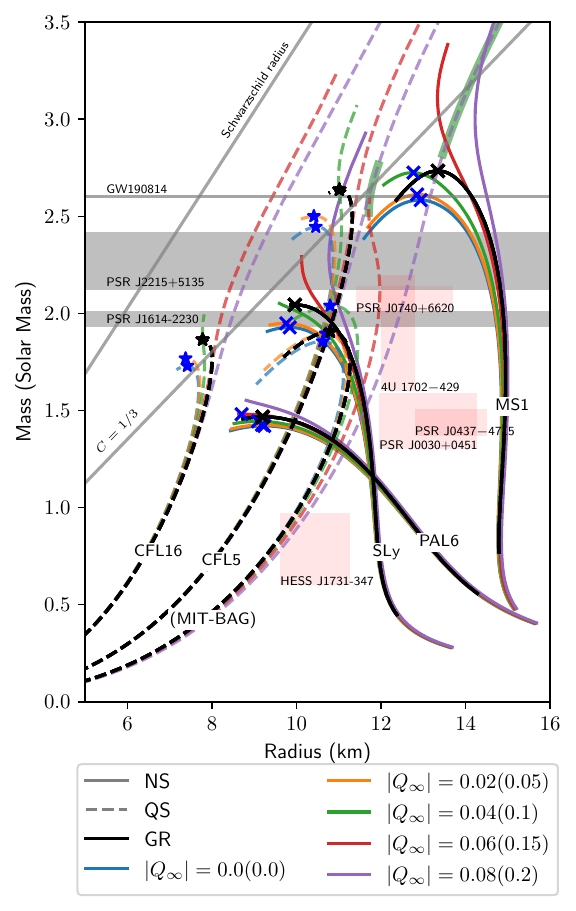}
	\caption[]{Mass and radius curve for $\eta=-1$ case of NS EOS from Ref. \cite{Read:2008iy} (solid lines) and QS EoS from CFL \cite{Lugones:2002va} and MIT-bag \cite{Farhi:1984qu} (dashed lines) for smaller values of $Q_{\infty}$. Salmon red color shows the measured M-R of pulsars from \cite{Miller:2019cac,Riley:2021pdl,Nattila:2017wtj,Gonzalez-Caniulef:2019wzi,Reardon:2015kba} and a central compact object of supernova remnant from \cite{Doroshenko}. Horizontal gray stripes show measured compact stars from \cite{LIGOScientific:2020zkf,Linares:2018ppq,Demorest2010}. Maximum mass marked with `x' for the NS and  `$\bigstar$' mark for the quark stars. Line colors represent the value of $Q_{\infty}$ where inside parentheses are values for (MIT-Bag). The black line shows the GR case for each EoS. Green shades show the condition where $F'^2$ is always positive inside the star.}
	\label{fig:mvsr_nsqs}
\end{figure}

\begin{figure}
	\centering
	\includegraphics[width=\linewidth]{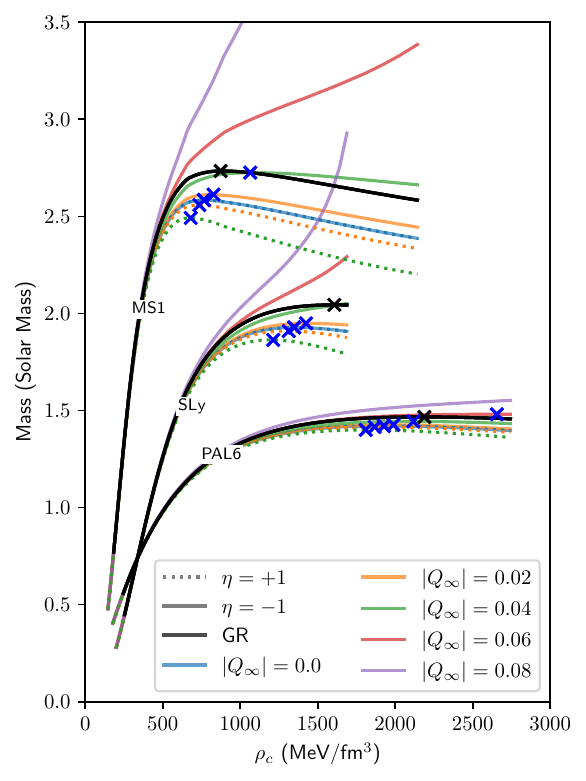}
	\caption[]{Mass-central density curve for NS EoS from Ref. \cite{Read:2008iy}. Solid line represents $\eta=-1$ case where dotted line represents $\eta=+1$. Black solid line represents GR case. Critical masses are labeled by `x'.}
	\label{fig:mvsrho_ns}
\end{figure}

 \textit{Numerical results.} It is interesting to discuss the mass-radius behavior at the small values of $|Q_{\infty}|$. The numerical solutions are shown in Fig. \ref{fig:mvsr_nsqs} for mass-radius (together with QS result) and \ref{fig:mvsrho_ns} for mass-central energy density at small $|Q_{\infty}|$. In contrast, Fig. \ref{fig:massrad_NS} shows mass and radius curves for large $|Q_{\infty}|$. Like the WD case, the $\eta=+1$ case tends to lower mass while $\eta=-1$ tends to increase it. The magnitude of mass discrepancy (relative to $|Q_{\infty}|=0$) is also controlled by the value of $|Q_{\infty}|$. However, in contrast to WD, the mass and radius modifications are more sensitive in high-density stars, while they reduce closer to GR in lower (central) density stars. These modifications are significant relative to GR above the order of hundredths of $|Q_{\infty}|$.
 
 Similar to the results from Ref. \cite{Cisterna:2015yla}, the $|Q_{\infty}| \rightarrow 0$ case does not revert to the GR. The small $|Q_{\infty}|$ has a slightly lower mass than GR in high-density regions. For the three NS EoSs we mentioned above, the critical mass of the GR case almost overlaps $|Q_{\infty}|=0.04$ in $\eta=-1$ case (see Fig. \ref{fig:mvsrho_ns}).
 
 For the exact value of $|Q_{\infty}|$ and $\eta$, it is apparent that the modifications are more prominent in a more stiff EoS. In some cases, $dM/d\rho_c$ is always positive, implying no maximum critical mass at these configurations. This occurs at $\eta=-1$ cases. It occurs at relatively low $|Q_{\infty}|$. For example, in SLy4 EoS, the maximum critical mass at the largest $|Q_{\infty}|$ -- where there are still existing turning points of mass-radius-central density -- is slightly lower than the critical mass of GR. This means that in some cases of EoS, the critical turning point of NS is consistently lower than that of GR, but it needs to get rid of the turning point to reach GR's critical mass.
 
 This also suggests that softer EoS can achieve higher mass, while stiffer EoS can simultaneously result in even greater mass. For example, SLy EoS could reach the mass of $2.6 M_\odot$, the mass of the secondary component of GW190814. In $|Q_{\infty}|>0.06$ cases on MS1 EoS, the mass could exceed $M>3.0 M_\odot$, reaching the range of stellar-mass BH. At the extreme central pressure in the above case, the mass-radius curve is approximately parallel to the Schwarzschild mass-radius, with the \textit{compactness} $\left( M/R \right)$ slowly approaching Schwarzschild's at larger $P_c$.
 
 \textit{Observational consequences.} Masses and radii are then compared with the observational measurement from X-ray pulsars, i.e., PSR J0030+0451 \cite{Miller:2019cac}, PSR J0740+6620 \cite{Riley:2021pdl}, 4U 1702-429 \cite{Nattila:2017wtj}, and PSR J0437-4715 \cite{Gonzalez-Caniulef:2019wzi,Reardon:2015kba}. Dynamical mass measured from massive binary stars also shown on Fig. \ref{fig:mvsr_nsqs}, i.e., secondary (less massive) component of GW binary GW190814 \cite{LIGOScientific:2020zkf}, PSR J2215+5135 \cite{Linares:2018ppq}, and J1614-2230 \cite{Demorest2010}. We exclude radius measurements derived from tidal deformability $\Lambda$ or measurements that depend on the structure equation to ensure that the measurement is independent of gravity theory.
 
 Figure \ref{fig:mvsr_nsqs} shows mass-radius for $\eta=-1$ case overlaid with observational mass-radius measurement data. At the range of about the canonical mass ($\sim 1.4M_\odot$), it requires the value of constant scalar at the order of $|Q_{\infty}|>0.1$ to modify the radius to match the observed data significantly. However, note that within this range, the choice of EoS also has a significant effect to fit with the data. For example, in the softer EoS, the mass and radius must be increased to fit with mass-radius data. Therefore, it needs $\eta=-1$. On the other hand, the MS1 EoS needs to reduce its mass and radius to fit the data in that range and, therefore, needs $\eta=+1$. 
 
 The more apparent signature could be inferred from the massive star. Note that the softer EoS (SLy4 and PAL6) cannot reach up to $2.6M_\odot$ (GW190814) without consequently removing the turning point of mass. This implies that if the EoS is soft, to predict massive NS within this theory, there would also be no 'mass gap' in NS distribution, as there is no critical mass due to large $|Q_{\infty}|$.
 
 \textit{Note on the scalar field in NS solution.} Despite the exciting properties of the $\eta=-1$ case, which could increase the critical mass of a star and predict high mass, the problem of negative $F'^2$ arises on all NS structures calculated in the $\eta=-1$ case. All these structures between $0<|Q_{\infty}|\leq0.08$ have negative $F'^2$ values around the center of the stars, where some of them always have the negative values of $F'^2$ inside the star for lower values of $|Q_{\infty}|$. However, this is not the case for $\eta=+1$, where $F'^2$ is always positive inside the NS structure.

\begin{figure}
	\centering
	\includegraphics[width=\linewidth]{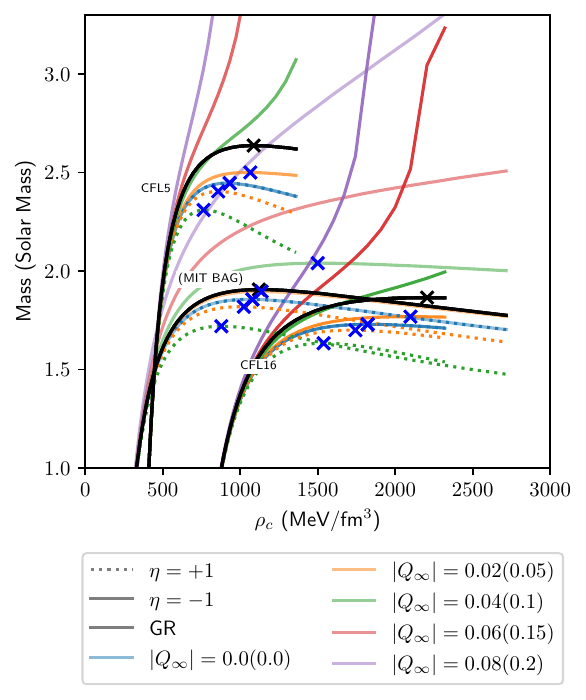}
	\caption[]{Similar to Fig. \ref{fig:mvsrho_ns} for quark star EOS. The value of $|Q_{\infty}|$ for MIT-Bag model represented inside the parentheses.}
	\label{fig:mvsrho_qs}
\end{figure}

\subsection{Quark star: MIT-bag and color-flavor locked EoS}
\textit{Quark star equation of state.} The potential existence of QSs is a consequence of the idea that the presence of strange quarks can lower the binding energy of strange quark matter (SQM) in weak equilibrium below that of $^{56}\rm Fe$ (absolute stability of quark matter). The MIT Bag model provides the most straightforward description for absolutely stable SQM \cite{Farhi:1984qu}. The quarks are free in that model, with confinement provided through a bag constant. However, the attractive force among anti-symmetric quarks in color tends to make quarks close to the Fermi surface paired at high densities. It has been shown that a color-flavor locked (CFL) state, in which quarks near the Fermi surface form pairs, seems to be more energetically favorable and widens the stability window~\cite{Lugones:2002va} (see also ~\cite{Paulucci:2008jd} and references therein). A recent review of the role of color superconductivity in dense quark matter can be found in Alford et al. ~\cite{Alford:2007xm}. It is also worth pointing out that a detailed analysis of pulsar timing data in pulsar evolution has shown that the SQM model is consistent with both radio and x-ray observations.
In contrast, the ordinary nuclear matter model requires enhancement by a dumping mechanism~\cite{Alford:2013pma}. CFL SQM can also be found in the inner cores of neutron stars (hybrid stars) ~\cite{Alford:2001zr}. Here, we use the simple MIT-Bag model \cite{Farhi:1984qu} and color-flavor locked (CFL) EoS \cite{Lugones:2002va} to represent strange stars EoSs.

The model EoS within the MIT-Bag model can be written as
\begin{align}
	P = \omega(\rho - 4B_0),
\end{align}
with $\omega$ is a parameter (dimensionless factor) related to the strange quark mass $m_s$ and coupling constant $\alpha_c$ in QCD \cite{Jaffe:1978bu}. For example, for $0 \leq m_s \leq 250$ MeV and $0 \leq \alpha_c \leq 0.6$, $\omega$ between 0.28 and 1/3 \cite{Zdunik:2000xx}. In this work, we choose the value of $\omega = 0.301$ and $B_0 = 56$ MeV/fm$^3$ which corresponds to $\alpha_c = 0.2$ and $m_s = 200$ MeV (or so-called "SQSB56" on Refs. \cite{GondekRosinska:2008zmv,GondekRosinska2007}).

The CFL EoS can be written in barotropic form as \cite{Lugones:2002va}
\begin{align}
	\rho = 3P+4B_0 - \frac{6\Delta^2 \mu^2}{\pi^2}+\frac{3m_s^2 \mu^2}{2\pi^2},
\end{align}
where
\begin{align}
	\mu^2 = \left[ \alpha^2 + \frac{4}{9}\pi^2(\rho-B_0) \right]^{1/2}-\alpha,
\end{align}
with
\begin{align}
	\alpha=\frac{2}{3}\Delta^2-\frac{m_s^2}{6}.
\end{align}
Here, we have three parameters ($B_0$, $m_s$, and $\Delta$) to be specified. There are multiple viable parameter set combinations that can represent CFL EoS. Here, we use parameter set from Ref. \cite{VasquezFlores:2017uor} and select the the lowest (CFL16, $M_{max} = 1.582 M_\odot$) and highest (CFL5, $M_{max} = 2.842 M_\odot$) maximum GR mass to represent this EoS. The CFL16 has $B_0 = 120$ MeV/fm$^3,$ $ m_s = 0$, and $\Delta=100$ MeV, while CFL5 has $B_0 = 60$ MeV/fm$^3$, $ m_s = 0$, and $\Delta=150$ MeV. The result of mass-radius calculations is shown in Fig. \ref{fig:mvsr_nsqs}.

\textit{Numerical result}. The calculated mass and radius from these EoSes can be seen on the dashed line of Fig. \ref{fig:mvsr_nsqs}. Like the NS case, the $|Q_{\infty}|\rightarrow0$ does not revert to GR. The modifications are also more prominent at massive stars, similar to the NS case, where $\eta=-1$ tends to increase the stellar mass. For CFL16 and CFL5 EoS, modification is significant above the value of hundredths, while for the MIT-Bag model, the modification is less sensitive and significant above the value of tenths.

Similar to the NS case, the condition with no maximum mass turning point is also present in the $\eta=-1$ scenario. The behavior resembles NS in higher central density, where the mass-radius curve parallels the Schwarzschild mass-radius. This scenario could reach GW190814 secondary component mass of $M=2.6M_\odot$ and higher up to stellar-mass blackhole range of $M>3.0M\odot$.

Like the NS case, the EoS choices are also significant within this range. Figure \ref{fig:mvsr_nsqs} is also overlaid by mass-radius measurement of the potential strange star, HESS J1731-347 \cite{Doroshenko}, which has relatively low mass and small size compared to pulsar measurements. The star coincides with MIT-Bag model in wide cases of $|Q_{\infty}|$ (including GR case), while for $|Q_{\infty}|\sim0.15$ to $|Q_{\infty}|\sim0.2$ in $\eta=-1$ case, MIT-Bag also coincide with PSR J0740+6620 and 4U 1702-429. The CFL16 and CFL EoS need larger radii to match with HESS J1731-347 and therefore need $|Q_{\infty}| \gtrsim 0.4$ with $\eta=-1$. Note that the quark stars are calculated without defining the existence of the crust. If the crust is considered, the overall radius would be slightly increased.

\textit{Note on the scalar field in QS solution.} Similar to NS, the $F'^2$ solution for $\eta=-1$ case also has a negative component in almost all calculated QS EoS, except the MIT-Bag model for $|Q_{\infty}|\gtrsim0.15$ at the very high central pressure (see green stripes at Fig. \ref{fig:mvsr_nsqs}). These have similar $F'^2$ characteristics with the NS case, where the negative $F'^2$ is more dominant in low $|Q_{\infty}|$. The structure of $F'^2$ on MIT-Bag is shown in Fig. \ref{fig:F2_bag}. Interestingly, at $|Q_{\infty}|\sim0.15$, the $2.6M_\odot$ MIT-Bag star has always positive $F'^2$ structure.

\begin{table}[h!]
	\caption{Threshold of $|Q_{\infty}|$ in which the stellar solution has critical mass limited by $dM/d\rho_c<0$ condition. The values above this threshold may virtually have no limiting critical mass $M_c$. Note that this only relevant for $\eta=-1$ case.}
	\label{tab:eosvsQthresh}
	\begin{ruledtabular}
		\begin{tabular}{lcc}
			EoS name & $|Q_{\infty}|$ Threshold & $M_{c,max}$ \\
			\hline
			MS1 & 0.048 & 2.837 \\
			SLy4 & 0.028 & 1.974 \\
			PAL6 & 0.060 & 1.480 \\
			MIT-Bag & 0.121 & 2.170 \\
			CFL16 & 0.022 & 1.779 \\
			CFL5 & 0.023 & 2.525 \\
		\end{tabular}
	\end{ruledtabular}
\end{table}

\textit{Existence of critical mass.} If we look more closely into the smaller values of $|Q_{\infty}|$, the threshold where there are no turning points in mass strongly depends on the EoS (see Table \ref{tab:eosvsQthresh}). For example, the MIT-Bag gives the largest threshold, MS1 gives the largest mass and radius (at the maximum critical point), and CFL16 gives the largest compactness. Interestingly, in SLy4, CFL16 and CFL5, the $M_{c,max}$ is slightly below its GR critical mass ($M_{c,GR}=2.04 M_\odot, 1.86 M_\odot$ and $2.64 M_\odot$ for SLy4, CFL16 and CFL5, respectively). In other words, in the above EoS, to recreate maximum mass in GR, the theory would also predict no critical mass due to (in)stability related to $dM/d\rho_c$. This situation might depend not only on the EoS stiffness but also on the overall mass-radius solution for each case.

The critical mass could be tested using the 'mass gap,' a gap between the most massive NS and the least massive black hole in the compact star population. In the above situation, to enable the massive star prediction (e.g., $M>2.5M_\odot$) by softer EoS would consequently remove the existence of the 'mass gap.' Therefore, the maximum values of $|Q_{\infty}|$ might be constrained by the upper limit of observed NS mass distribution. The current upper limit is $\sim 2.6 M_\odot$, assuming the secondary component of GW190814 is a compact star. With this value, even with this modified gravity, only stiff EoS with $M_{crit, GR}\gtrsim 2.5 M_\odot$ would be able to explain the gap. However, this does not consider another constraint that might determine the gap even with the nonexistent critical mass, such as the stability of the matter structure or the stability of the scalar field.

\subsection{Stability issue of the scalar field}
\begin{figure}
	\centering
	\includegraphics[width=\linewidth]{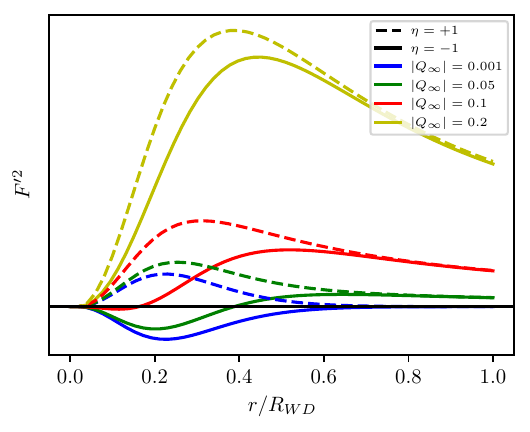}
	\caption[]{The structure of $F'^2$ inside WD with $P_c = 4\times 10^{-4}$ MeV fm$^{-3}$ (around critical mass) for various $|Q_{\infty}|$. The positive $\eta$ cases are shown by dashed line, while negative $\eta$ shown by solid lines. Colors (except black) represents different value of $|Q_{\infty}|$. Horizontal solid black line indicate $F'^2 = 0$. All have a similar radius of about $R\sim500$ km}
	\label{fig:F2_WD}
\end{figure}

\begin{figure}
	\centering
	\includegraphics[width=\linewidth]{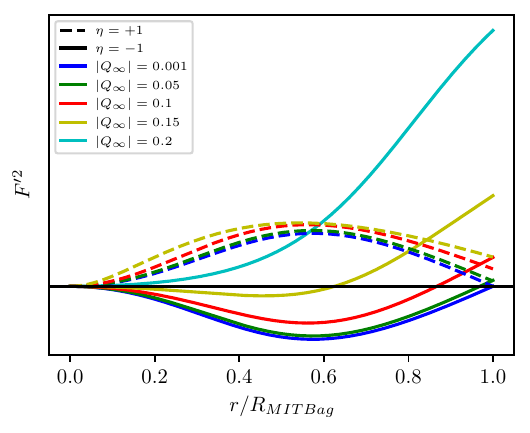}
	\caption[]{The structure of $F'^2$ inside MIT-Bag star with $P_c = 500$ MeV fm$^{-3}$ depicted with a legend similar to that in Fig. \ref{fig:F2_WD}. Note that the stars calculated here have significant radius differences. }
	\label{fig:F2_bag}
\end{figure}

The behavior of the scalar field in neutron star solution depends on the behavior of function $F'(r)$ and has been discussed briefly in Ref. \cite{Cisterna:2015yla}. The authors notice that imaginary valued $F'(r)$ arise when the sign of $\eta$ is negative. This fact, in turn, may lead to classical and quantum instabilities, but these authors have not yet identified whether this is true. Then, in Ref. \cite{Cisterna:2016vdx}, the authors discuss this problem of negative $\eta$ much further. There, the authors state that their paper's negative $\eta$ results cannot be trusted because of the possibility of quantum instabilities. However, the result shows an increase in the mass of the compact star, which is also what we obtained and reported in this paper. Moreover, the authors also mention that the mass increases as the central density increases without any turning point when $\eta$ is negative. Therefore, no maximum mass is observed for the neutron star models they investigate. Our results show a phenomenon similar to that of both the neutron star and quark star models but not of the white dwarf model. Furthermore, we also check the behavior of the scalar field by plotting $F'(r)^2$ in Figs.~\ref{fig:F2_WD} and \ref{fig:F2_bag} for white dwarfs and MIT Bag stars, respectively. Both results are from picking a single central pressure while we vary the values of $\eta$ and $|Q_\infty|$. The solid lines are negative $\eta$, and the dashed lines are positive $\eta$. Here, we focus only on the value of $F(r)$ if it is real or complex. As shown in Figs.~\ref{fig:F2_WD} and \ref{fig:F2_bag}, $F(r)$ is complex-valued for the majority of negative $\eta$ except when $|Q_\infty|=0.2$. We can see this if we observe Eq.~\eqref{eq:Fprimesquaredcenter} when we use negative $\eta$:
	\begin{equation}
		F'(r_c)^2 = \left( -\frac{P_c}{|\eta|} + \frac{2Q^2(3P_c-\rho_c)}{3(3Q^2|\eta|+4\kappa b_c^2)} \right) r_c^2.
	\end{equation}
The first term inside the bracket can be suppressed momentarily because this is valid only when $r\simeq r_c$, when $Q^2$ is sufficiently large, producing a larger $|Q_\infty|$. This is the reason why $F'(r)^2>0$ if $|Q_\infty|$ is large enough. Note that when $\eta=+1$, we see a decrease in the maximum mass. This condition is valid regardless of the equation of state that we use. Moreover, The scalar field function $F'$ is real on all $r$. 

Interestingly, no complete stellar solution is obtained when $|Q_\infty|$ is too large in $\eta=+1$. This is primarily caused by the increasing pressure from the stellar center to its surface instead of decreasing due to the positive pressure gradient. From Eqs. \eqref{eq:dpdr} and \eqref{eq:dbdr}, $P'(r)>0$ means $b'(r)<0$ which imply $f(r)>1$ at some $r$. If we look at Eq. \eqref{eq:f'} and look into the stellar center, we have
\begin{equation}
	f'(r_c)=\frac{2 b_c r_c \left(3 P_c+\rho _c\right)}{3 Q^2 \left| \eta \right| -b_c \left(P_c r_c^2+4 \kappa \right)}>0.
\end{equation}
Therefore, in these cases, $f(r_c)$ would increase from the one outward to the center.

Besides the conditions of the scalar field solution, another instability might arise from a hairy solution prone to Laplacian instability \cite{Kase2021}. This behavior may occur in the case of compactness $C = GM/R<1/3$ via even-parity perturbation related to the scalar-field propagation along the angular direction at the surface. Therefore, in our case, this instability may occur at almost all EoS solutions, except the massive part of NS or QS (see $C=1/3$ line on Fig. \ref{fig:mvsr_nsqs}).

\section{Concluding remark}
\label{conclu}

The stellar structure solution of the non-minimal derivative coupling term of Hordenski's gravity predicts the different stellar mass and radius properties from general relativity. We investigated the white dwarf, neutron star, and quark star properties predicted by the theory in two cases, $\eta = +1$ and $\eta=-1$. We assume a static, spherically symmetric configuration of the star. We aim to see more closely the prediction of the compact stars' maximum mass among various equation of states and to check their compatibility with the relevant observational data.

It has been noticed before that stellar masses and radii in $|Q_{\infty}| \rightarrow 0$ case are not reduced to GR in NS (although both are close to GR). Similar behavior in both NS and QS EoSs is found in this study. Interestingly, in the WD case, the stellar properties (i.e., mass, radius, energy, and pressure structures) coincide with GR at $|Q_{\infty}| \rightarrow 0$. This result indicates that in the $|Q_{\infty}| \rightarrow 0$ case, the gravity tends to approach GR (or Newtonian) in low density, non-relativistic scale. It should be noted that the relativistic and non-relativistic structure calculation will give a slightly different value of critical mass.

Additionally, positive or negative $\eta$ affects the change in mass and radius of compact stars, with the magnitude dependent on $|Q_{\infty}|$. Generally, the negative $\eta$ tends to increase the maximum mass of the compact stars, while the positive $\eta$ tends to decrease them. The larger $|Q_{\infty}|$, the stronger the increase/decrease effect.

In WD, the critical masses fall at approximately a similar radius in GR. This fact enables us to constrain the minimum critical mass via massive WD pulsar data. This gives $|Q_{\infty}|\lesssim0.4$ as the strong upper bound for $\eta=-1$. The maximum radius concerning the rotational stability from the observed spin period of RX J0648.0-4418 also yields an upper bound for the $\eta=-1$ case with $|Q_{\infty}| \lesssim 0.6$. The measurement uncertainties of ultracold WD also give a constraint at lower mass, with $|Q_{\infty}|\lesssim0.2$ for both $\eta=\pm 1$ cases. Additionally, we note that even though the large value of $|Q_{\infty}|$ enables the prediction of very massive WD (above the super-Chandrasekhar mass of $M\sim 2.0M_\odot$), in the lower density, the corresponding $|Q_{\infty}|$ values are incompatible with the majority of WD measurements. Therefore, the theory alone may not be sufficient to explain the existence of super-Chandrasekhar supernovae.

However, that is not the case for NS and QS EoSs, where the mass-radius curves are more sensitive to $|Q_{\infty}|$ at higher density. For example, in the negative $\eta$ case with a relatively small value of $|Q_{\infty}|$, the high-density mass-radius of NS and QS could significantly increase while retaining close to GR in lower density. We can also set a threshold of $|Q_{\infty}|$ where the mass will always increase with the increase of central pressure (i.e., $dM/d\rho_c$ is always positive). In this case, the maximum turning point of mass (critical mass) as the upper bound does not exist. The above threshold is located in the relatively low value of $|Q_{\infty}|$, where in some cases of EoS, the critical mass only exists below GR critical mass.

Despite the occurrence of numbers of instability in $\eta=-1$, it is interesting to examine it from a phenomenological perspective. The above situation where $dM/d\rho_c$ is always positive is also found in degenerate higher-order scalar-tensor theory (DHOST) \cite{Kobayashi2018} along with significant alteration at higher density. On the other hand, critical mass could be tested by the maximum NS mass measurement below the lower mass gap, the mass between NS and blackhole. In the current knowledge, it is generally assumed that GW190814 secondary star mass is the highest NS mass. However, the existence of this value of mass gap is still an ongoing debate (see discussions on Refs. \cite{Lopes2022,Biswas2021} and recent report on Ref. \cite{KAGRA:2021duu}) and therefore need further measurements to construct mass population around this gap. For example, the future generation of gravitational wave detectors \cite{Abbott2020} would enable more sensitive detection of binary coalescence events.
In contrast, space-based (e.g., \cite{Sesana2016, Luo2015}) may detect binary systems at lower orbital frequency, far before coalescence. These observations would significantly increase the number of measured compact object mass. The similar mass distribution could also measured from x-ray binary \cite{Shao:2022qws}. If the above observation can provide a clear mass gap with higher significance, the critical mass of a compact star could be determined as the upper bound to rule out the value of $|Q_{\infty}|$.

\begin{acknowledgments}
I. Prasteyo is grateful for the Internal Research Grant No. 003/IRG/SU/AY.2023-2024 from CETL Sampoerna University. B. E. Gunara and A. Suroso acknowledge ITB Research Grant for financial support. 
\end{acknowledgments}

\appendix

\section{Appendix}

In this Appendix, we provide a mass-radius for a considerable value of $|Q_{\infty}|$ compared to GR results and the interior-exterior match condition of all objects considered in this work to test the correctness of the calculations.

\begin{figure}
	\centering
    \vspace*{2.5\baselineskip}
	\includegraphics[width=\linewidth]{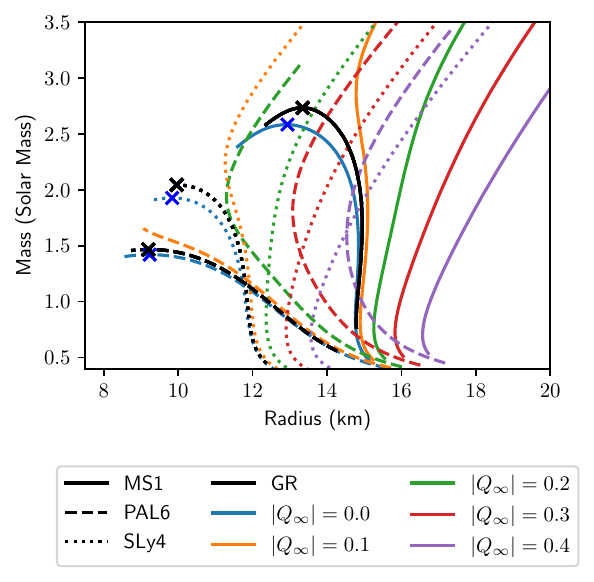}
	\caption[]{Mass and radius of NS EoSs for larger value of $|Q_{\infty}|$.}
	\label{fig:massrad_NS}
\end{figure}

\begin{figure}
	\centering
	\includegraphics[width=\linewidth]{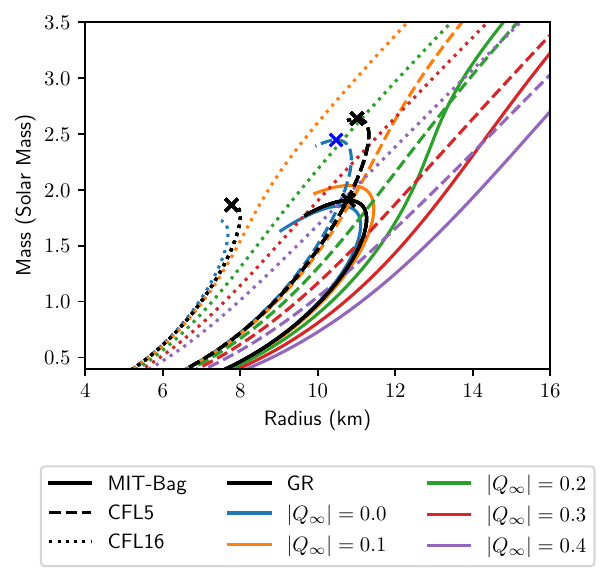}
	\caption[]{Mass and radius of QS EoSs for larger value of $|Q_{\infty}|$.}
	\label{fig:massrad_QS}
\end{figure}

\begin{figure}
	\centering
	\includegraphics[width=\linewidth]{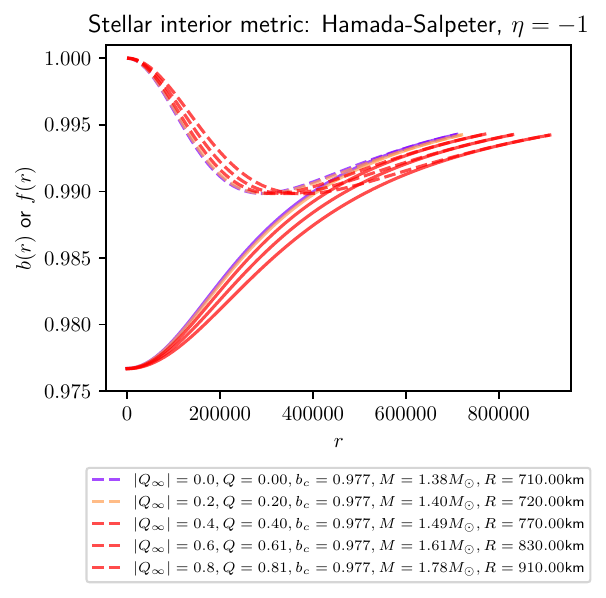}
	\caption[]{Internal metric solutions of Hamada-Salpeter WD with $P_c = 10^{-4}$ MeV fm$^{-3}$. Solid lines and dased lines represent $b(r)$ and $f(r)$, respectively.   }
	\label{fig:radvsbfWD}
\end{figure}

\begin{figure}
	\centering
	\includegraphics[width=\linewidth]{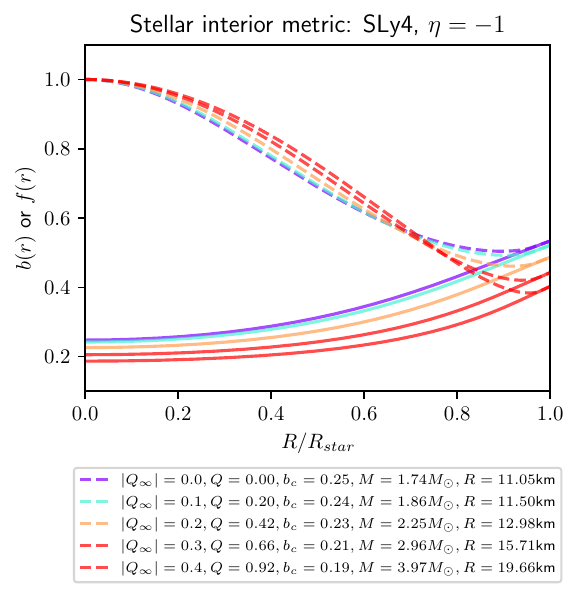}
	\caption[]{Internal metric solutions of SLy4 EoS with $P_c = 200$ MeV fm$^{-3}$. Solid lines and dased lines represent $b(r)$ and $f(r)$, respectively. }
	\label{fig:radvsbfSLy4}
\end{figure}

\begin{figure}
	\centering
	\includegraphics[width=\linewidth]{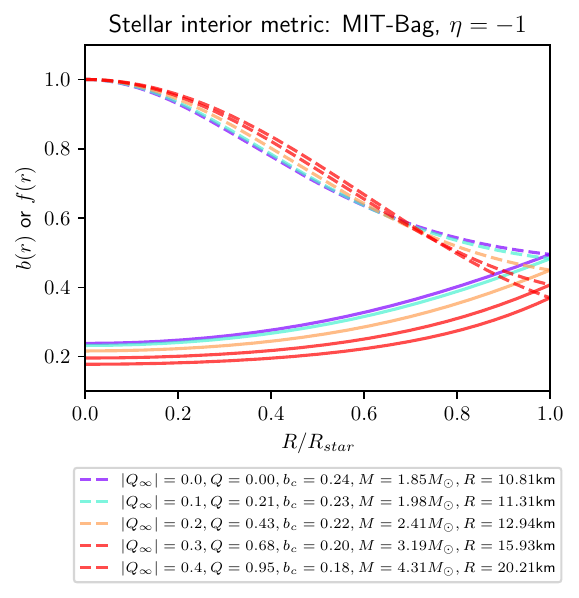}
	\caption[]{Internal metric solutions of MIT-Bag EoS with $P_c = 200$ MeV fm$^{-3}$. Solid lines and dased lines represent $b(r)$ and $f(r)$, respectively.}
	\label{fig:radvsbfBag}
\end{figure}

\bibliography{ref.bib}

\end{document}